\documentclass[12pt,preprint]{aastex}

\newcommand{\Vec}[1]{\mbox{\boldmath $#1$}}

\begin{document}


\shorttitle{Dissipation of Magnetic Flux in Primordial Clouds}
\shortauthors{Maki \& Susa}

\title{Dissipation of Magnetic Flux in Primordial Gas Clouds}

\author{Hideki Maki and Hajime Susa}
\affil{Department of Physics, Rikkyo University,
       Nishi-Ikebukuro, Tokyo 171-8501}
\email{maki@hel.rikkyo.ne.jp, susa@rikkyo.ac.jp}


\begin{abstract}
We report the strength of seed magnetic flux of accretion disk
surrounding the PopIII stars. The magnetic field in accretion disk
might play an important role in the transport of angular momentum
because of the turbulence induced by Magneto-Rotational
Instability (MRI). On the other hand, since the primordial
star-forming clouds contain no heavy elements and grains,
they experience much different thermal history and different
dissipation history of the magnetic field in the course of
their gravitational contraction, from those in the
present-day star-forming molecular clouds. In order to assess
the magnetic field strength in the accretion disk of PopIII stars, we
calculate the thermal history of the primordial collapsing clouds,
and investigate the coupling of magnetic field with
primordial gas. As a result, we find that the magnetic field
strongly couple with primordial gas cloud throughout
the collapse, i.e. the magnetic field are frozen to the gas as far as
initial field strength satisfies
$ B\lesssim 10^{-5}(n_{\rm H}/10^3~{\rm cm^{-3}})^{0.55}~{\rm G}$.
\end{abstract}



\section{INTRODUCTION}
PopIII stars are considered to have very important impacts on the thermal
history and the chemical evolution of the universe. 
Recent observations  of the polarization of Cosmic Microwave Background
(CMB) photons by WMAP(Wilkinson Microwave Anisotropy Probe)
\citep{Kogut03} revealed that the optical depth of the universe by Thomson
scattering is fairly large ($\tau = 0.17 \pm 0.04$). This result tells
that reionization epoch is earlier ($z = 17 \pm 5$) than expected from
the observations on Gunn-Peterson trough \citep{Becker01,Fan02,White03}. 
On the other hand, the results of theoretical calculations of structure
formation \citep{Sokasian03} indicate that
such early reionization seems to be difficult solely by the Pop II
stars, but it might be possible if the ionizing photons from PopIII
stars with top heavy initial mass function (IMF) is taken into account. Thus, the mass (or IMF)
of PopIII star have great significance on the reionization epoch of the
universe. In addition, some amount of metals are found at high redshift
intergalactic matter by the observation of high redshift QSO absorption
line systems\citep{songaila01,ma02,vladilo02}. This means significant
fraction of baryons are already processed in the stars by $z=5$ \citep{songaila01}.
PopIII stars also should be
responsible for such metal pollution of the early universe, and the
abundance pattern of heavy elements depends on the mass of PopIII stars.
Thus, typical mass (or IMF) of the PopIII stars is a key quantity also for the
chemical evolution of the universe.

In order to assess the typical mass of PopIII stars, typical scale of the
prestellar core (or fragments of primordial gas)
should be evaluated as a first step. 
Several authors have studied on this issue by simple one-zone
approach before middle of '90s \citep{MST69,Hut,Carl,PSS83,SUN96,USNY96,Puy}. 
Recently, multi-dimensional numerical simulations of fragmentation of
primordial gas have been performed intensively by several
authors\citep{nakamura99,bromm99,abel00,nakamura01,bromm02}, and they
find that the mass of PopIII prestellar core is quite massive ($\sim
10^{3-4}M_\odot$). 

Further evolution of prestellar core was investigated by
\cite{omukai98}, and it is found that the collapse proceeds in a run-away
fashion and converges to Larson-Penston similarity solution
\citep{Larson69,Penston69} with $\gamma\simeq 1.1$.
They also find the mass accretion rate is also very
large compared to the present-day forming stars, although spherical
symmetry is assumed in their numerical calculation.
In reality, however, prestellar cores have some amount of angular
momentum, which prevent the mass from accreting onto the protostars.
Consequently, accretion disks surrounding the protostars are expected, and
some mechanisms that transport their angular momentum are required in order
to enable the mass accretion onto protostars.

There are a few possibilities of the angular momentum transport
such as gravitational torque by the nonaxisymmetric structures in the
accretion disk, the interaction among the 
fragments of the disk \citep{stone00,bodenheimer00}, 
and the turbulent viscosity
triggered by Magneto-Rotational Instability (MRI)
\citep{hawley92,sano98,sano01}.
All of these mechanisms are regarded to
be important for present-day star formation.
Thus, it is worth to investigate these processes for primordial case.
Among these possibilities, we concentrate on the last one, the MRI
induced turbulence.

In order to activate MRI, the initial magnetic field strength in the
accretion disk should be larger than a critical value. Otherwise the
magnetic field is dissipated before the field is amplified
by MRI \citep{TM04,TB04}.
Thus, it is important to assess the ``initial'' magnetic field
strength in the accretion disk. As was discussed by
\citet{nakano86} for
preset-day case, magnetic field could be dissipated while
prestellar core collapses. However, the dissipation processes strongly
depend on the components of the gas as well as the the temperature
evolution in the course of the collapse. On the other hand, the
temperature of the collapsing primordial gas is much higher than that
of the present-day case \citep{omukai00}, which might bring about
different dissipation history of the magnetic field.

In this paper, we investigate the dissipation of magnetic field in
collapsing primordial gas cloud. Consequently, we obtain the initial
magnetic field strength in the accretion disk of PopIII stars. In the
next section, formulations are described. In
\S\ref{results}, results of our calculations are shown, and the
importance of MRI in PopIII star formation is discussed
in \S\ref{discussion}. Final section is
devoted to summary.

\section{CALCULATIONS}
\label{calculations}
We consider the collapse of spherical clouds without metals and
dusts, but with slight magnetic fields. Then, we calculate
the time evolution of the central density, chemical composition
and dissipation of magnetic flux from the gravitationally collapsing
core. The initial strength of magnetic fields when primordial
clouds begin to contract is given as unknown parameter.

\subsection{Dynamics}
We assume that the dynamics is described by the free-fall relation,
\begin{equation}
	\frac{d \rho}{dt}=\frac{\rho}{\tau_{\rm ff}},
\end{equation}
where $\rho$ is the density of the collapsing core and
$\tau_{\rm ff} = (3 \pi/32G\rho)^{1/2}$ is the free-fall time.
In fact, as described in \citet{omukai00}, thermal
evolution of the core in 1D hydrodynamic simulation is well described by
such one zone model. 
On the other hand, we also have to evaluate the dissipation of
magnetic field in the accretion phase, and it is not the same as the
case of collapsing core. In this case, the temperature of the accretion flow
would rise faster than those in the core, because of shock
heating. Thus, the magnetic field is expected to be frozen to the gas also in
the accretion flow, if the frozen condition is satisfied in the
collapsing core. 

It is also worth noting that this free-fall approximation is also
marginally valid for the case of quasi-hydrostatic core found in the
simulation by \citet{abel00,abel02}. 
In fact, panel C of Fig.3 in \citet{abel00} tells that
the cooling time is as short as the free-fall time. Thus, the system
evolves with time scale $\tau_{\rm ff}\simeq \tau_{\rm cool}$.

In order to describe thermal processes, we approximate the equation of
state to be polytrope with index $\gamma=1.1$,
\begin{equation}
	p=K \rho^{\gamma},
\end{equation}
where $K$ is the constant coefficient. The reason to adopt
the polytrope of $\gamma=1.1$ is that in past work the thermal
evolution of primordial collapsing core is investigated in detail
by \citet{omukai00}, where the temperature grows with
$\gamma \sim 1.1$  in considering heating/cooling processes.

Since we are interested in the collapsing gas cloud,
the magnetic energy density needs to be less than
the gravitational energy density of the cloud core.
The critical field strength $B_{\rm cr}$ is defined by the equation
$B_{\rm cr}^2/4\pi R \simeq \rho GM/R^2$. Since the radius of the core
should be on the order of Jean's length, $B_{\rm cr}$ is given by the
following equation,
\begin{eqnarray}
	B_{\rm cr} &\simeq& (3G)^{1/2}M_{\rm J}/R_{\rm J}^2 \nonumber \\
    	       &=& (4/3)^{1/2}\pi^{3/2}c_{\rm s}\rho^{1/2}\propto \rho^{0.55},
\label{eq:critical_field}
\end{eqnarray}
where $R_{\rm J}$, $M_{\rm J}$, and $c_{\rm s}$ are the
Jean's length, the Jean's mass, and the sound velocity,
respectively.

\subsection{Dissipation of Magnetic Fields}
There are two processes for the dissipation of magnetic
fields.\footnote{Another process is the diffusion of unidirectional
magnetic fields by turbulent flows \citep{kim02},
which is beyond the scope of this paper,
since we investigate the frozen-in condition of the directional fields. }
The first one is Ohmic loss, and the other is ambipolar diffusion.
We assess the degree of dissipation defined by the ratio of the drift
velocity $v_{{\rm B}x}$ of the field lines from the gas
to the free-fall velocity, as was introduced in \citet{nakano86}.
In \citet{nakano86}, the drift
velocity involves both the Ohmic dissipation and the ambipolar
diffusion processes. There are two important quantities which characterize
these diffusion processes. They are $\tau_\nu$ and $\omega_\nu$ which denote
the viscous damping time of the relative velocity of charged particle $\nu$
to the neutral particles,  and the cyclotron frequency of
the charged particle $\nu$, respectively. 
Then,  $\tau_\nu$ is expressed as
\begin{equation}
	\tau_\nu = \frac{\rho_\nu}{\mu_{\rm \nu n}
				n_\nu n_{\rm n}\langle\sigma v\rangle_{\rm \nu n}},
\end{equation}
where $\mu_{\rm \nu n}$ is the reduced mass, $n_\nu$, $n_{\rm n}$,
and $\rho_\nu$ are, the mean number density for
the charged particle $\nu$, the neutral particle ${\rm n}$, and the mass
density of charged particle $\nu$, respectively.
The averaged  momentum-transfer rate coefficient for a
particle $\nu$ colliding with a neutral particle is expressed by
$\langle \sigma v \rangle_{\rm \nu n}$ .

The momentum-transfer cross section of an ion with
a neutral particle is given by the Langevin rate coefficient
\citep{osterbrock61}. For an electron, it is found experimentally
by \citet{hayashi81} that the cross sections at low energies are
much smaller than the Langevin rate coefficient
and are nearly equal to a geometrical cross section.
We use the empirical formulae for the momentum-transfer rate coefficients
\citep{kamaya00,sano00}.

According to \citet{nakano86}, the drift velocity
can be given by
\begin{equation}
	v_{{\rm B}x} = \frac{A_1}{A}\frac{1}{c}(\Vec{j}\times\Vec{B})_x,
\label{eq:diffusion_velocity}
\end{equation}
where
\begin{eqnarray}
	A = A_1^2 + A_2^2, \label{eq:A}\\
	A_1 = \sum_\nu \frac{\rho_\nu \tau_\nu \omega_\nu^2}
		{1 + \tau_\nu^2 \omega_\nu^2}, \label{eq:A1}\\
	A_2 = \sum_\nu \frac{\rho_\nu \omega_\nu}
		{1 + \tau_\nu^2 \omega_\nu^2}, \label{eq:A2}
\end{eqnarray}
$\Vec{B}$ is the mean magnetic field in the primordial cloud,
the suffix $x$ means $x$ direction component in local Cartesian
coordinates where the $z$ direction is taken as the direction of
$\Vec{B}$. In calculating the drift velocity $v_{{\rm B}x}$ according to equation
(\ref{eq:diffusion_velocity}) we replace
$(1/c)(\Vec{j} \times \Vec{B})_x$ with the mean magnetic force
$B^2/4 \pi R$, where $B$ is the mean field strength in the cloud,
$R$ is the radius of the cloud.

The magnetic field is mainly dissipated by the Ohmic loss 
when $|\tau_\nu \omega_\nu| < 1$ for main charged particles.
In such a situation we have the approximate expression from
equations (\ref{eq:diffusion_velocity}), (\ref{eq:A}),
(\ref{eq:A1}) and (\ref{eq:A2}) that
\begin{equation}
	v_{{\rm B}x} \sim \frac{c^2}{4 \pi \sigma_{\rm c} R},
\end{equation}
where $\sigma_{\rm c}$ the electrical conductivity,
\begin{equation}
	\sigma_{\rm c} = \sum_\nu \frac{q_\nu^2\tau_\nu n_\nu}{m_\nu},
\end{equation}
$q_\nu$ and $m_\nu$ are the electrical charge and the mass for
a charged particle $\nu$, respectively. Thus the drift
velocity is independent of $B$. On the other hand, when
$|\tau_\nu \omega_\nu| > 1$, the ambipolar diffusion is
a main process of dissipation.
In such a case we obtain
\begin{equation}
	v_{{\rm B}x} \sim \frac{\tau_{\rm i}}{\rho_{\rm i}}\frac{B^2}{4\pi R},
\end{equation}
where the suffix i expresses the ion particles.
Note that the drift velocity is proportional to $B^2$ in this case.

\citet{TB04} have evaluated the drift velocity of magnetic field in the
quasi-hydrostatic core found in \citet{abel00,abel02}, when its
density is a certain value. However, this system is
highly non-equilibrium, and the ionization degree gets
smaller and smaller as the collapse proceeds. Thus, its is
not trivial at all whether the frozen condition is satisfied
or not at much higher density. Therefore, we have to
perform detailed non-equilibrium calculations of chemical
reaction network,
in order to obtain the correct ionization degree in the collapsing gas.

\subsection{Chemistry}
In order to investigate the evolution of ionized fraction
during the collapse in detail, we solve non-equilibrium
chemical reaction network of primordial gas
that involves not only H element, but also D, He,
and Li. Furthermore, we introduce following 24 species:
${\rm e^-}$, ${\rm H^+}$, H,
${\rm H^-}$, ${\rm H_2}$, ${\rm H_2^+}$, ${\rm H_3^+}$, D,
${\rm D^+}$, $\rm D^-$, HD, ${\rm HD^+}$, ${\rm H_2D^+}$,
He, ${\rm He^+}$, ${\rm He^{++}}$, ${\rm HeH^+}$, Li,
${\rm Li^+}$, ${\rm Li^{++}}$, ${\rm Li^{3+}}$, ${\rm Li^-}$,
LiH, and ${\rm LiH^+}$.
We employ the latest reaction rate coefficients appearing in the
following papers, \citet{galli98}, \citet{omukai00},
\citet{stancil98}, \citet{flower02} and \citet{lepp02}.
As for the radiative recombination, we use the rate coefficients based on
\citet{spitzer78}.

\subsection{Initial Conditions}
We employ two different initial conditions A and B.
For model A, we set $n_{\rm H,ini}=10^3\ {\rm cm^{-3}}$
and $T_{\rm ini}=250\ {\rm K}$
for initial conditions of prestellar core. We use the fraction of
the chemical composition in the early universe given by \citet{galli98} as the
initial condition for the cloud. This model is on the evolutionary track
($Z=0$, fiducial model) in \citet{omukai00}. On the other hand, as was
shown by \citet{PSS83} and \citet{omukai00}, the paths
in $n_{\rm H}-T$ plane rapidly converge to an identical trajectory from
various
initial conditions. Remark the almost hydrostatic prestellar core found
in the numerical simulations in \citet{abel00} is also very close to the
above trajectory.
Thus, this choice of initial condition is quite
natural for the formation of PopIII stars.

On the other hand, there is a case which do not converges rapidly to
the trajectory, which was pointed out by \citet{uehara00} and
\citet{nakamura02}. In this case, the gas is initially heated up
to 10$^4$ K during the collapse of rather massive host galaxy,
and H$_2$ is formed efficiently utilizing the residual electrons as
catalysts (e.g. Susa et al. 1998). Afterwards, HD is formed from H$_2$, and
the gas is cooled down to $\sim $100 K.
Furthermore, the gas temperature is kept at
relatively low level ($\sim 100$ K) in the course of the collapse. We
employ this case as model B, in which we set the initial
condition ($n_{\rm H,ini}=10^3\ {\rm cm^{-3}},
T_{\rm ini}=100\ {\rm K}$), and the initial fractional
abundances are taken from the results of \citet{nakamura02}.

The initial magnetic field strength for the primordial
star-forming environment has been investigated by many authors
(e.g. Pudritz \& Silk 1989; Kulsrud et al. 1997; Widrow 2002;
Langer, Puget \& Aghanim 2003).
However, the initial field generation is still a controversial
question.  Thus, the initial
magnetic field strength is given as the model parameter
in our calculations.

\section{RESULTS}
\label{results}
\subsection{Ionization degree}
\begin{figure}
\epsscale{0.85}
\plotone{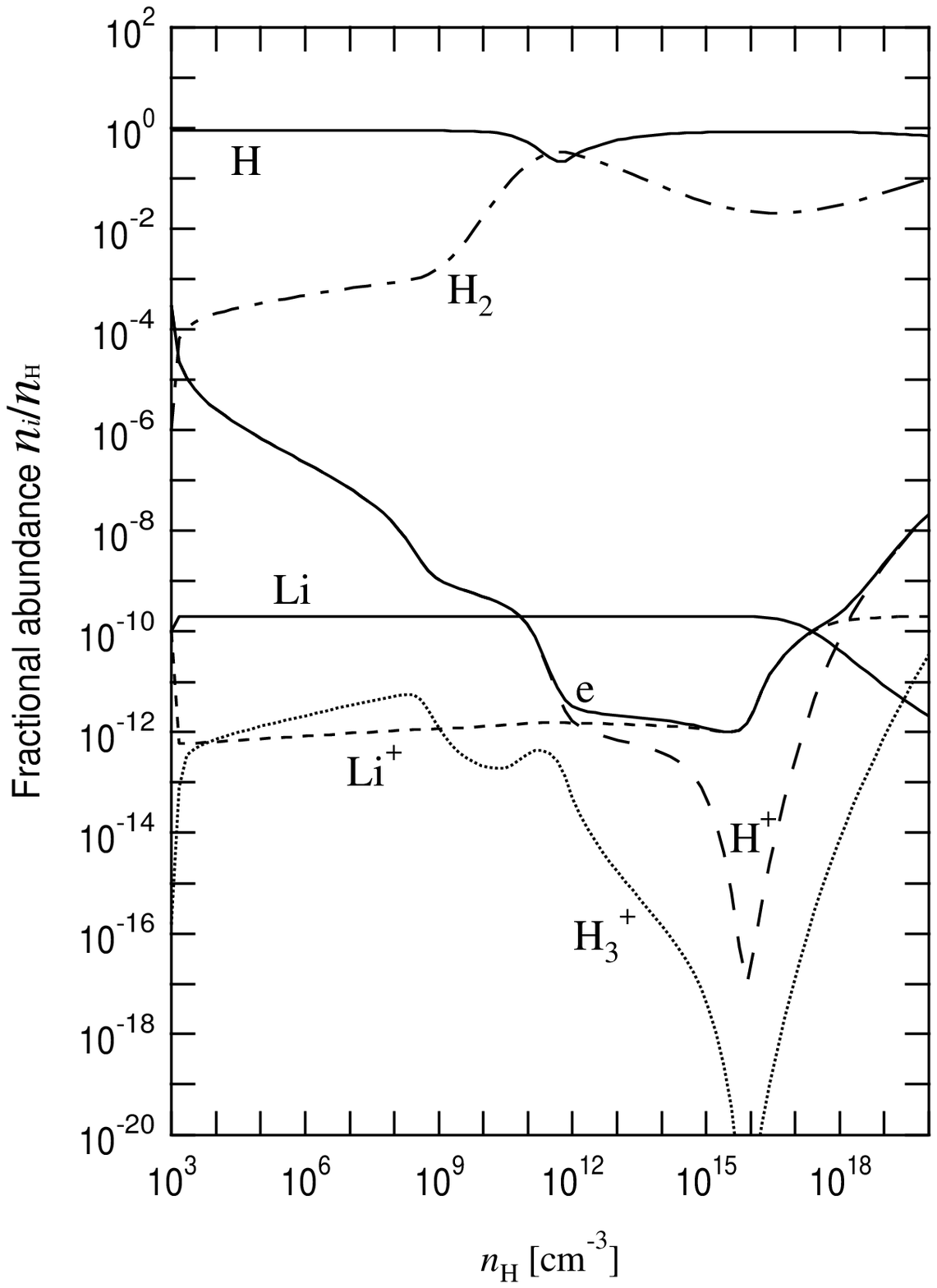}
\caption[dummy]{The evolution of fractional abundance of 
main species, e, ${\rm H}^+$, H, ${\rm H}_2$, ${\rm H_3^+}$, Li, and ${\rm Li}^+$,
are plotted for model A. Vertical axis denotes the fractional abundance of
the above each species and horizontal axis is the density of
the cloud.}
\label{fig:chem_A}
\end{figure}

\begin{figure}
\epsscale{1}
\plotone{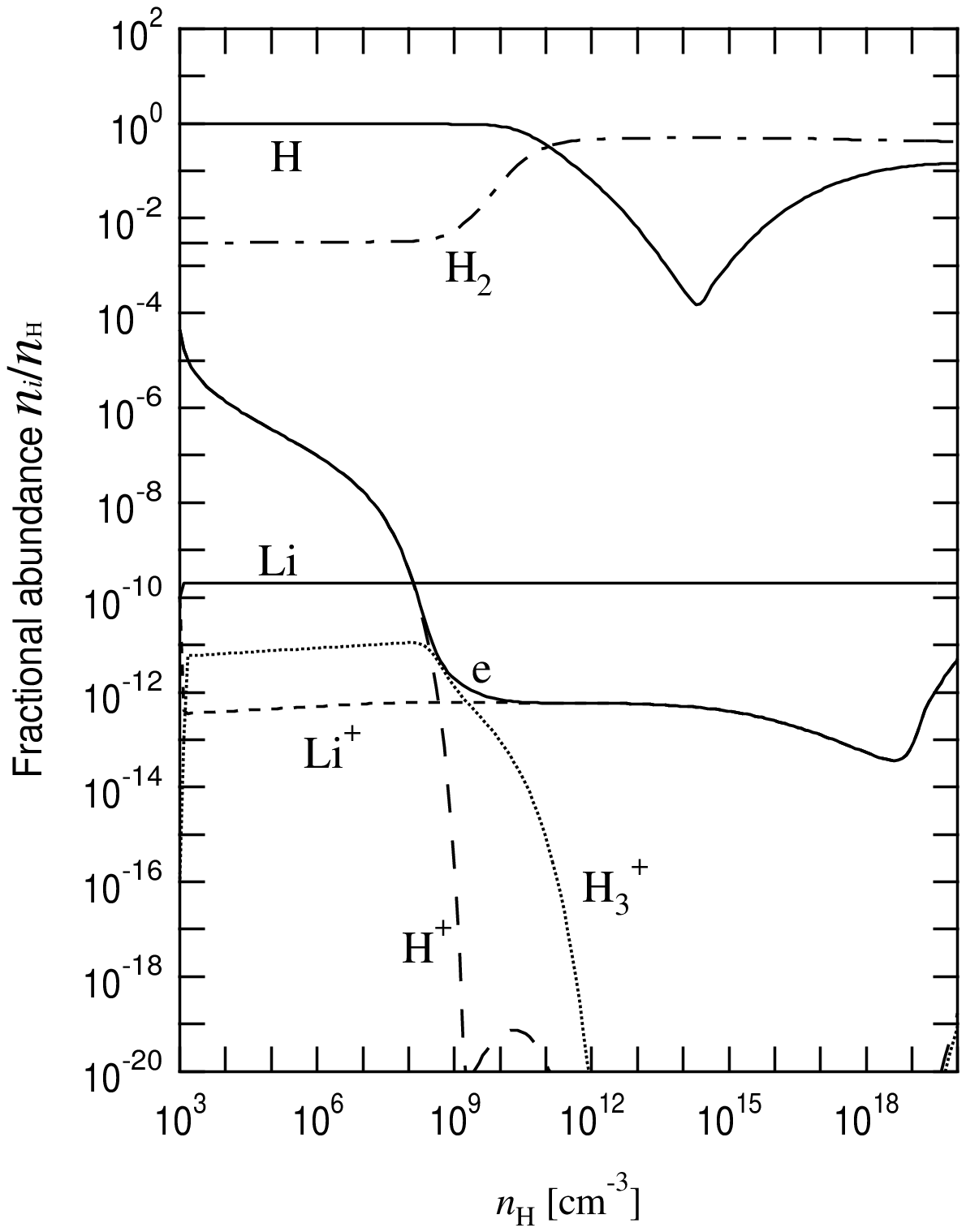}
\caption[dummy]{Same as Figure \ref{fig:chem_A} except that the initial
 temperature is 100K (model B).}
\label{fig:chem_B}
\end{figure}

Figure \ref{fig:chem_A} shows the fractional abundance of
various species, e, ${\rm H}^+$, H, ${\rm H}_2$, ${\rm H_3^+}$,
Li, and ${\rm Li}^+$ for model A.
At low density, $n_{\rm H} \lesssim 10^{11}\ {\rm cm}^{-3}$,
the ionization degree decreases as density increases
because of the recombination of
the electrons with the protons. However, since the ${\rm Li}^+$
recombination rate is smaller than that of ${\rm H}^+$,
the reduction of the ionization degree is almost quenched at 
$n_{\rm H}\sim 10^{12}\ {\rm cm^{-3}}$.
Moreover, the ionization degree increases at the higher density
because the collisional ionization is activated by high temperature.
Note that the ionization degree
is very small, but there is a minimal value $\sim 10^{-12}$.
Figure \ref{fig:chem_B} shows the results for model B. In this case, the
fraction of electrons evolve in a similar way. There is a minimal value
$\sim 10^{-12}$, at which the electrons are provided by Li.

\begin{figure}
\plotone{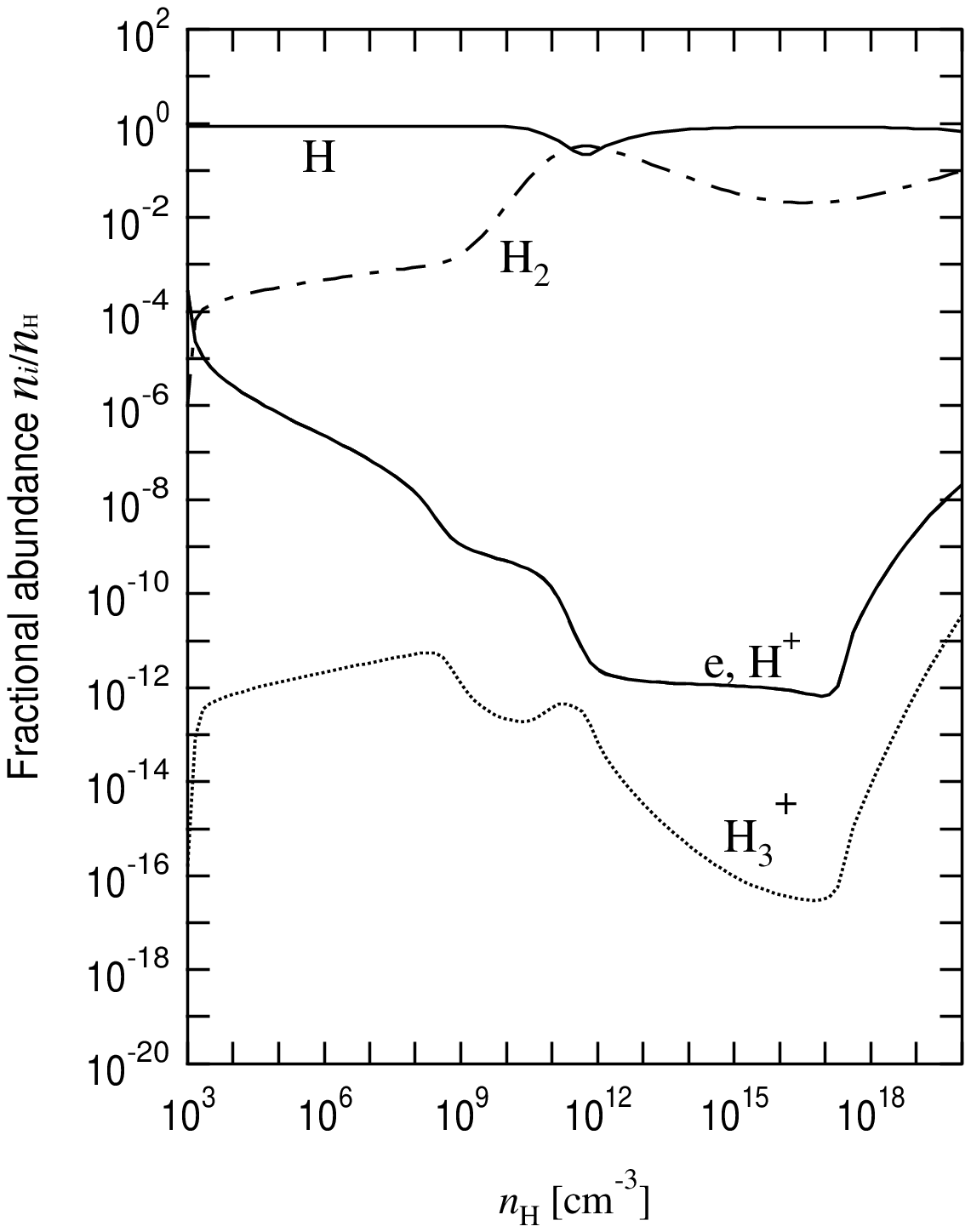}
\caption[dummy]{Same as Figure \ref{fig:chem_A} except that this model is
 calculated without Li.}
\label{fig:chem_Ac}
\end{figure}

\begin{figure}
\plotone{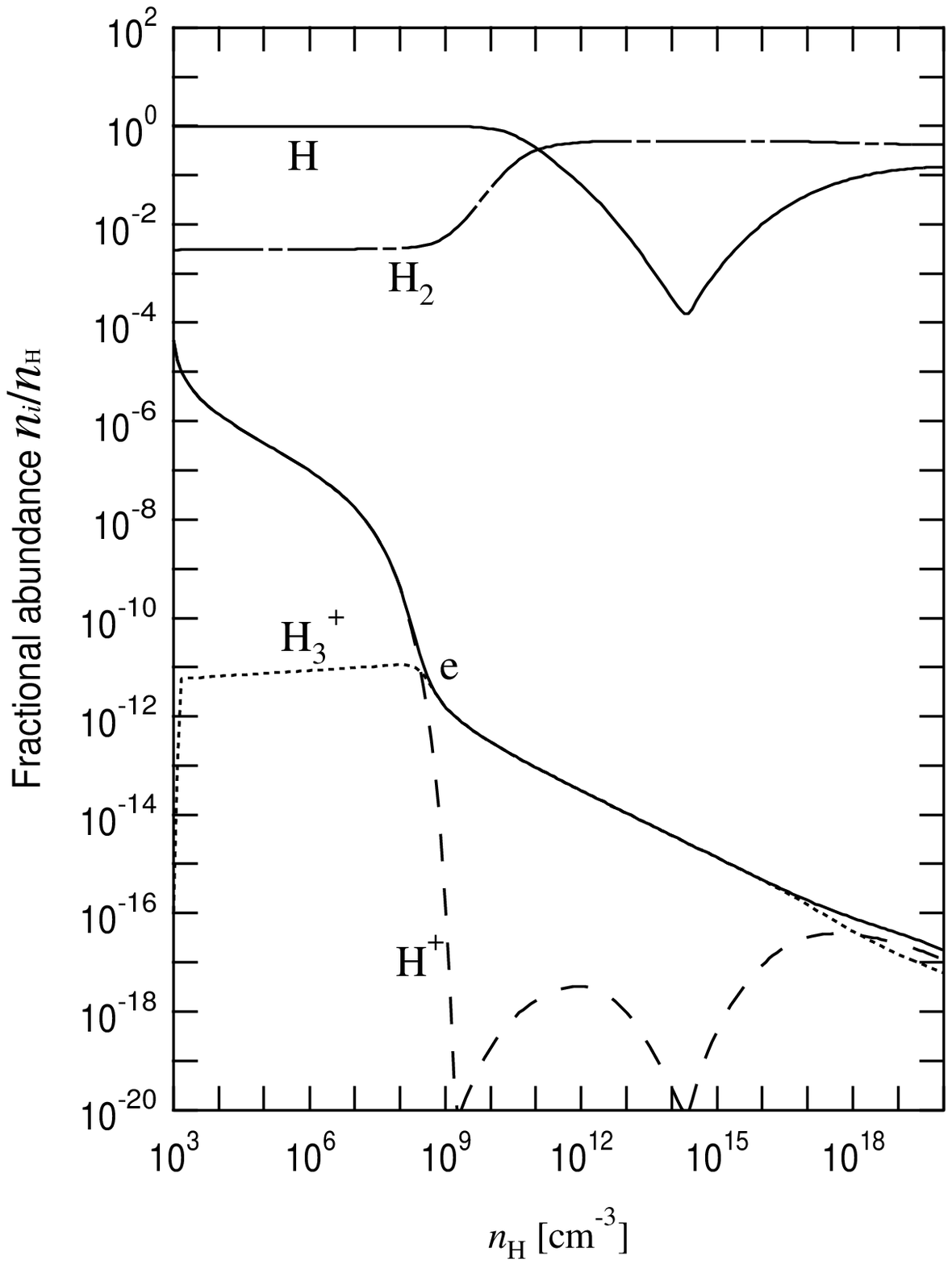}
\caption[dummy]{Same as Figure \ref{fig:chem_B} except that this model is
 calculated without Li.}
\label{fig:chem_Bc}
\end{figure}

In order to clarify the importance of Li, we perform control runs
without Li, ${\rm Li^+}$, ${\rm Li^{++}}$, ${\rm Li^{3+}}$, ${\rm Li^-}$,
LiH, and ${\rm LiH^+}$, for both initial conditions A and B. The results
of these runs are shown in Figures \ref{fig:chem_Ac} and
\ref{fig:chem_Bc}. 
In the case of model A, even
without Li, ionized fraction do not gets smaller than $\sim
10^{-12}$(Figure \ref{fig:chem_Ac}). Therefore, presence of Li is not so
important in this case. 
However, for the model B, 
the result is much different from the other case. 
In the presence of Li, electron fraction is kept around $10^{-12}$
even at $n_{\rm H} > 10^{11}\ {\rm cm^{-3}}$.
But in the absence of Li, electron
fraction gets much smaller than $10^{-12}$ (Figure \ref{fig:chem_Bc}). Thus,
Li is quite important in the low temperature model. We will come back to
this point in the next subsection.

\begin{figure}
\plotone{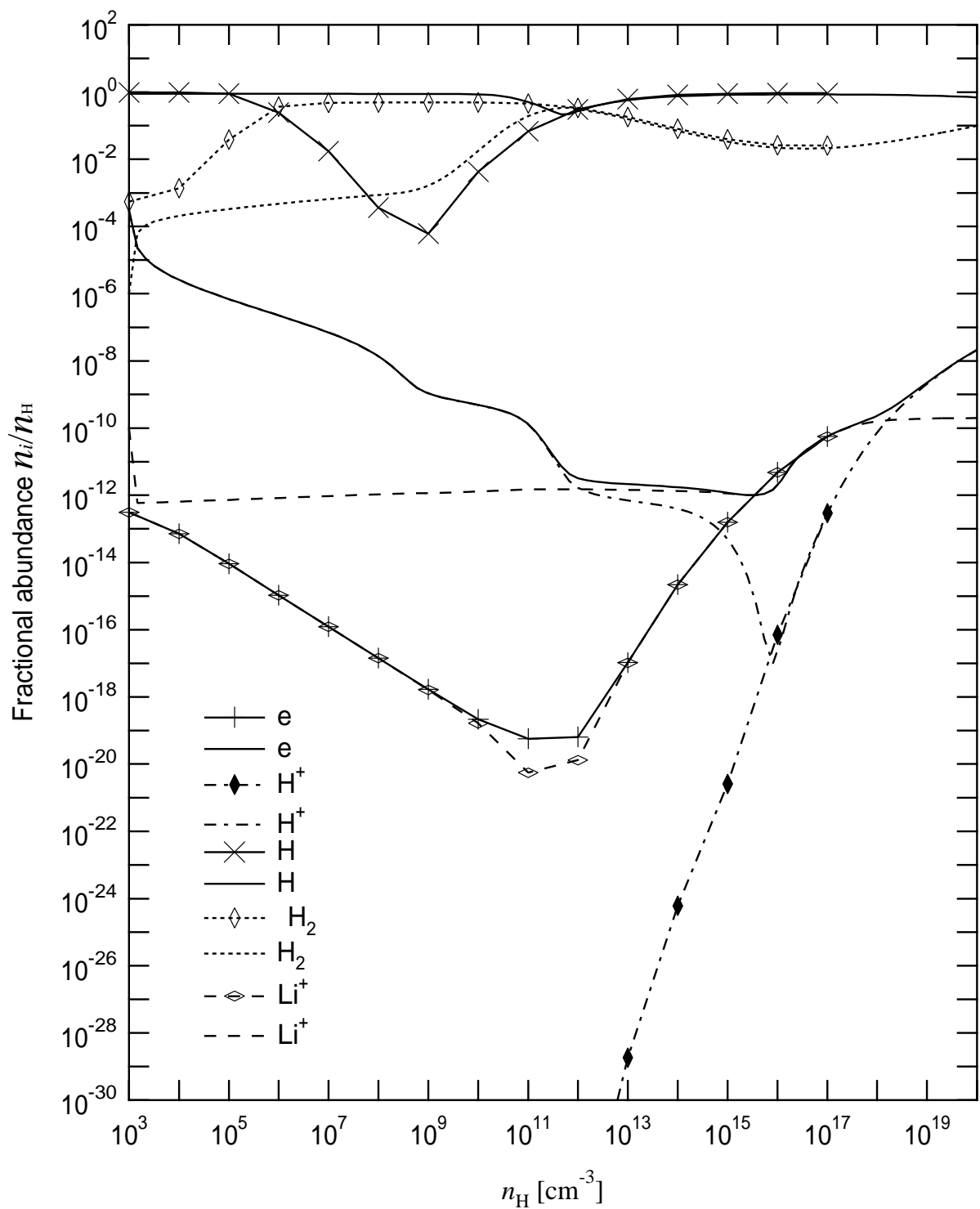}
\caption[dummy]{
The fractions of various chemical species
after the time integration assuming fixed density and temperature
over $10^3\ {\rm Gyrs}$ are superimposed on Figure \ref{fig:chem_A}.
Curves with markers represent the fractions after long time integration,
 and the unmarked curves represent the non-equilibrium fractions.
}
\label{fig:equi}
\end{figure}

\begin{figure}
\plotone{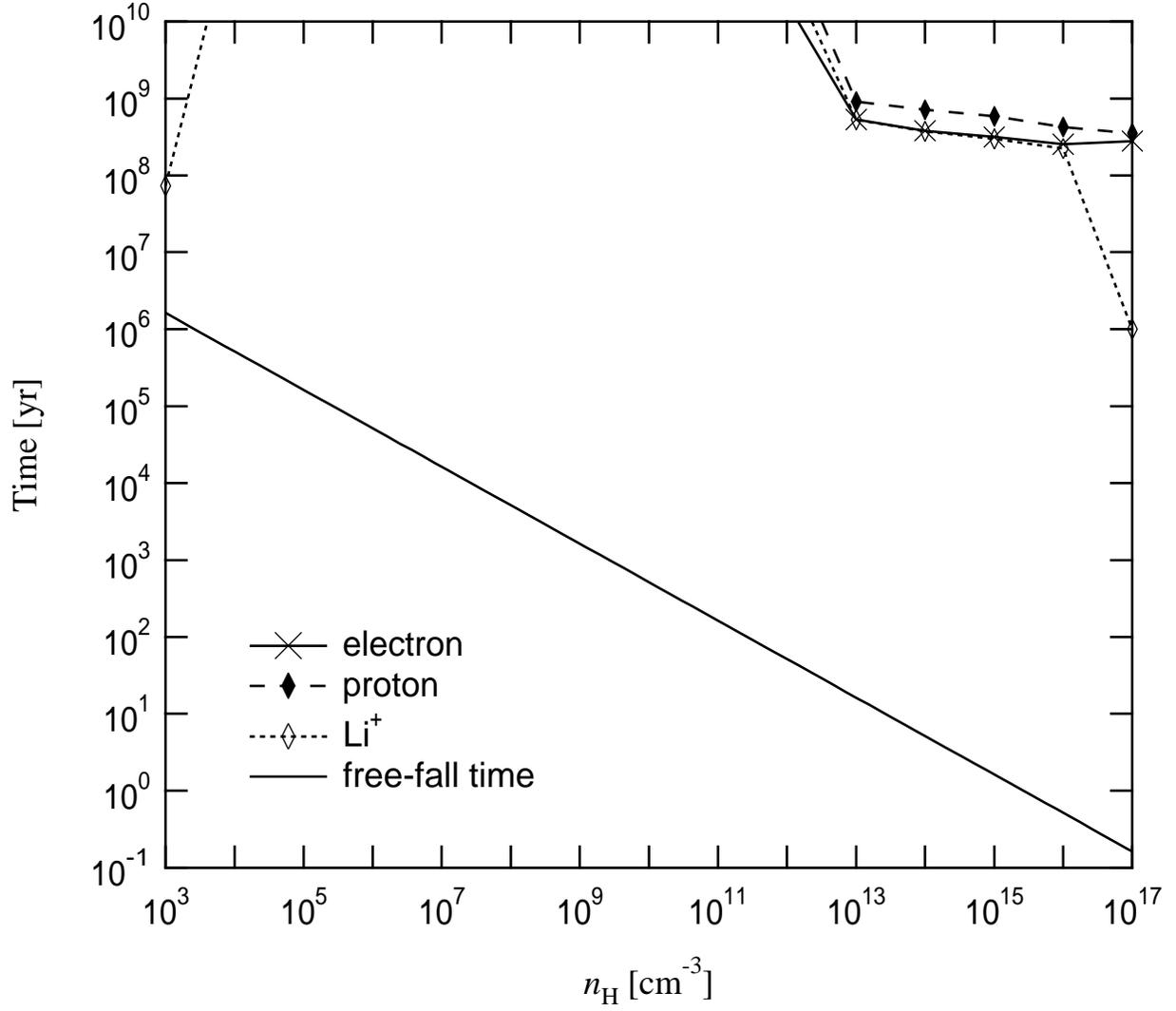}
\caption[dummy]{Time scale of equilibration is shown for several species.}
\label{fig:eqtime}
\end{figure}

We also emphasize the significance of non-equilibrium treatment of
chemical reactions. 
Figure \ref{fig:equi} shows the fractions of various chemical species
after the time integration assuming fixed density and temperature
over $10^3\ {\rm Gyrs}$ (a hundred times larger than the age of
the universe). As a result, for $n_{\rm H} \gtrsim 10^{13}\ {\rm cm^{-3}}$,
the fractions of chemical species converge to equilibrium values,
although they are still not in equilibrium for lower densities.
It is clear that results from non-equilibrium calculation is
very different from those in chemical equilibrium. Especially, the
non-equilibrium fraction of electron is much larger
than the equilibrium value. 
The time scales of equilibration are also shown in figure \ref{fig:eqtime},
and they are always much longer than the local free-fall time.
Thus, non-equilibrium calculations are indispensable to evaluate the
ionization degree of collapsing primordial gas.

\subsection{Drift velocity of magnetic field}
\begin{figure}
\plotone{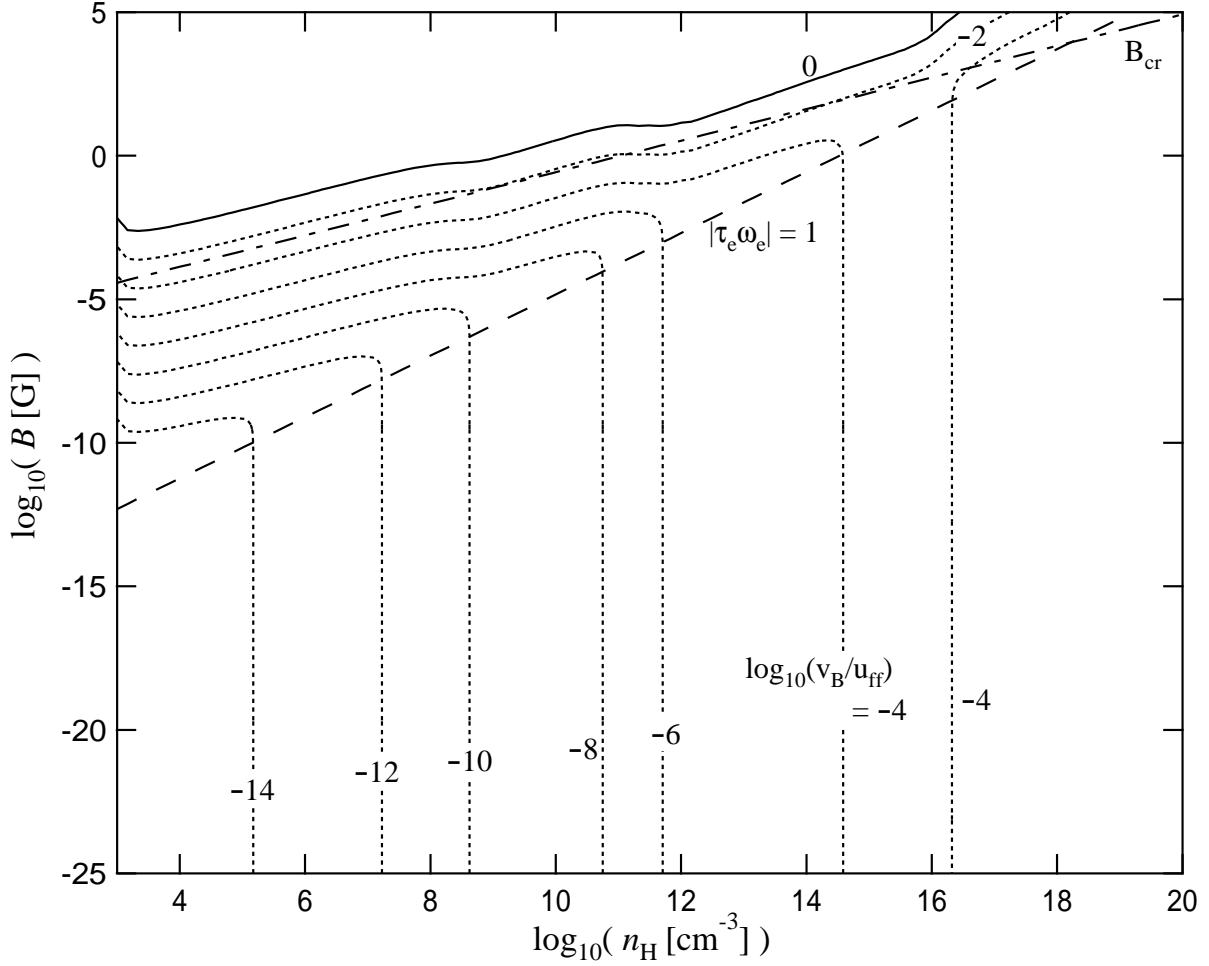}
\caption[dummy]{The drift velocity $v_{{\rm B}x}$ of the magnetic
field as a function of the density $n_{\rm H}$ and the
field strength $B$ for a collapsing cloud for model A. The dotted curves
represent the contours of constant $v_{{\rm B}x}$ normalized
by the free-fall velocity $u_{\rm ff}$ and its logarithmic values are
labelled on those curves.
The solid curve shows the locus
along which $v_{{\rm B}x}=u_{\rm ff}$ is satisfied.
The dashed line represents the locus $|\tau_\nu \omega_\nu|=1$
for electrons. The dot-dashed line represents
the critical field $B_{\rm cr}$ given by equation
(\ref{eq:critical_field}).}
\label{fig:dv_A}
\end{figure}

\begin{figure}
\plotone{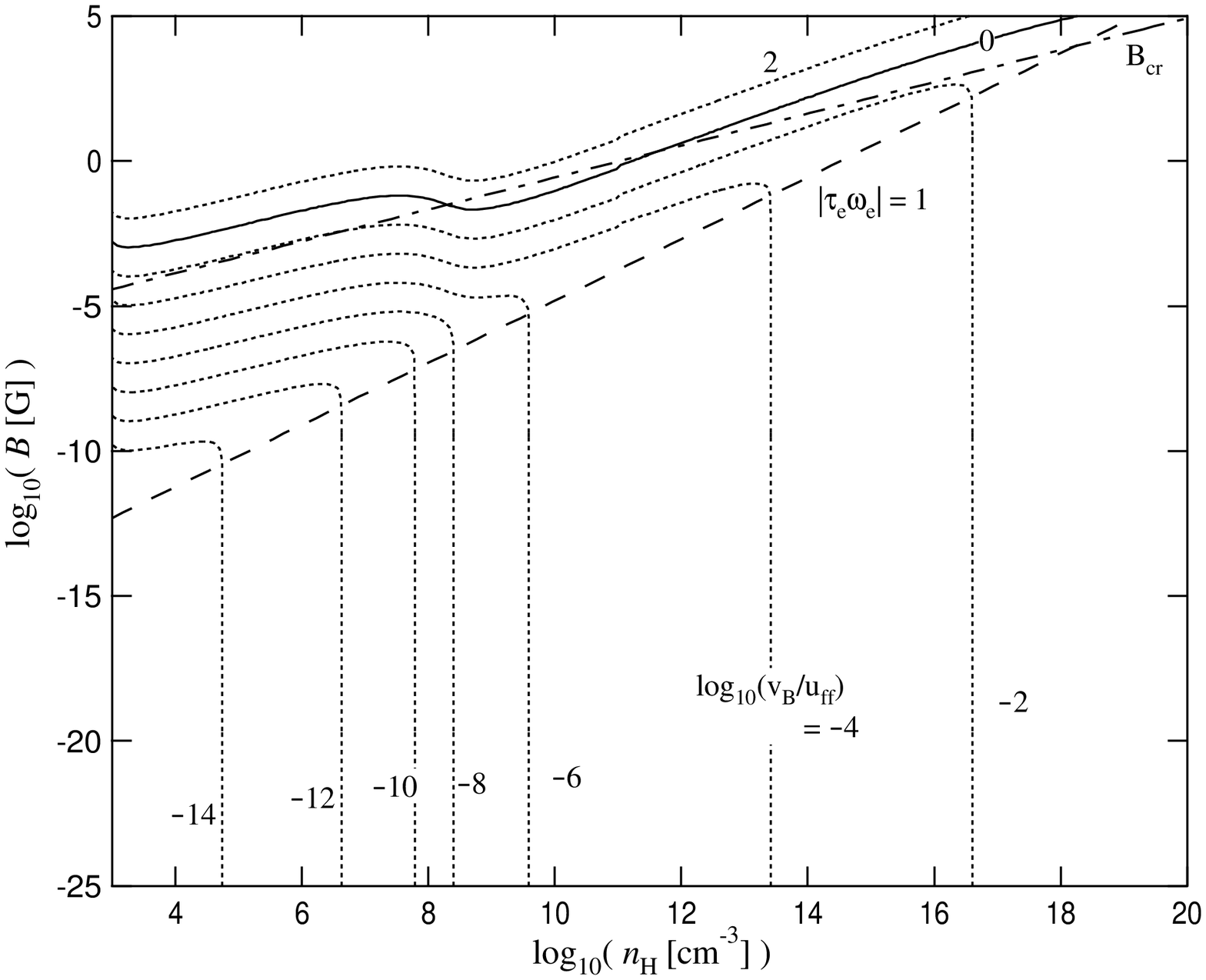}
\caption[dummy]{Same as Figure \ref{fig:dv_A} except that the initial
 temperature is 100K (model B).}
\label{fig:dv_B}
\end{figure}

\begin{figure}
\plotone{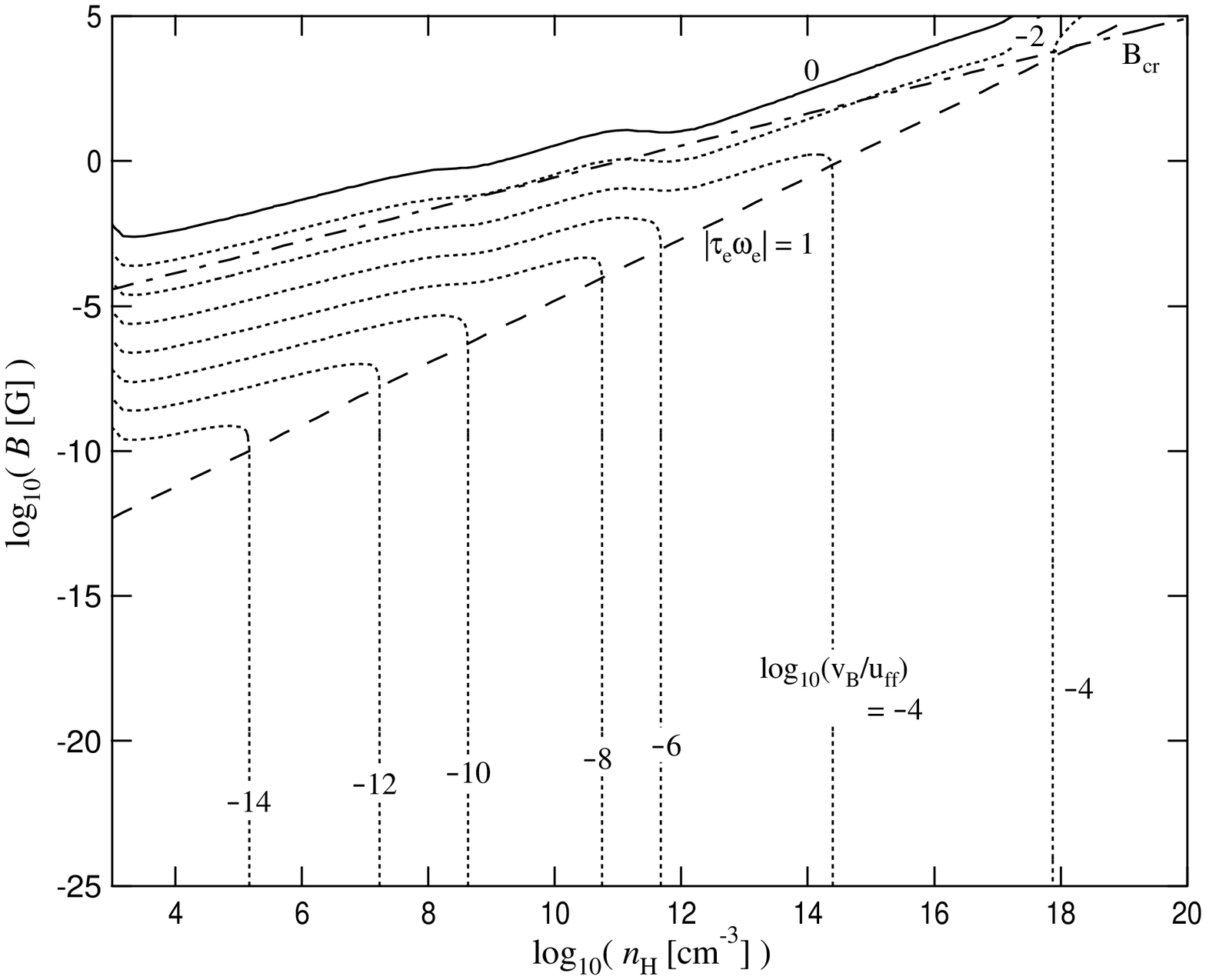}
\caption[dummy]{Same as Figure \ref{fig:dv_A} except that this model is
 calculated without Li.}
\label{fig:dv_Ac}
\end{figure}

\begin{figure}
\plotone{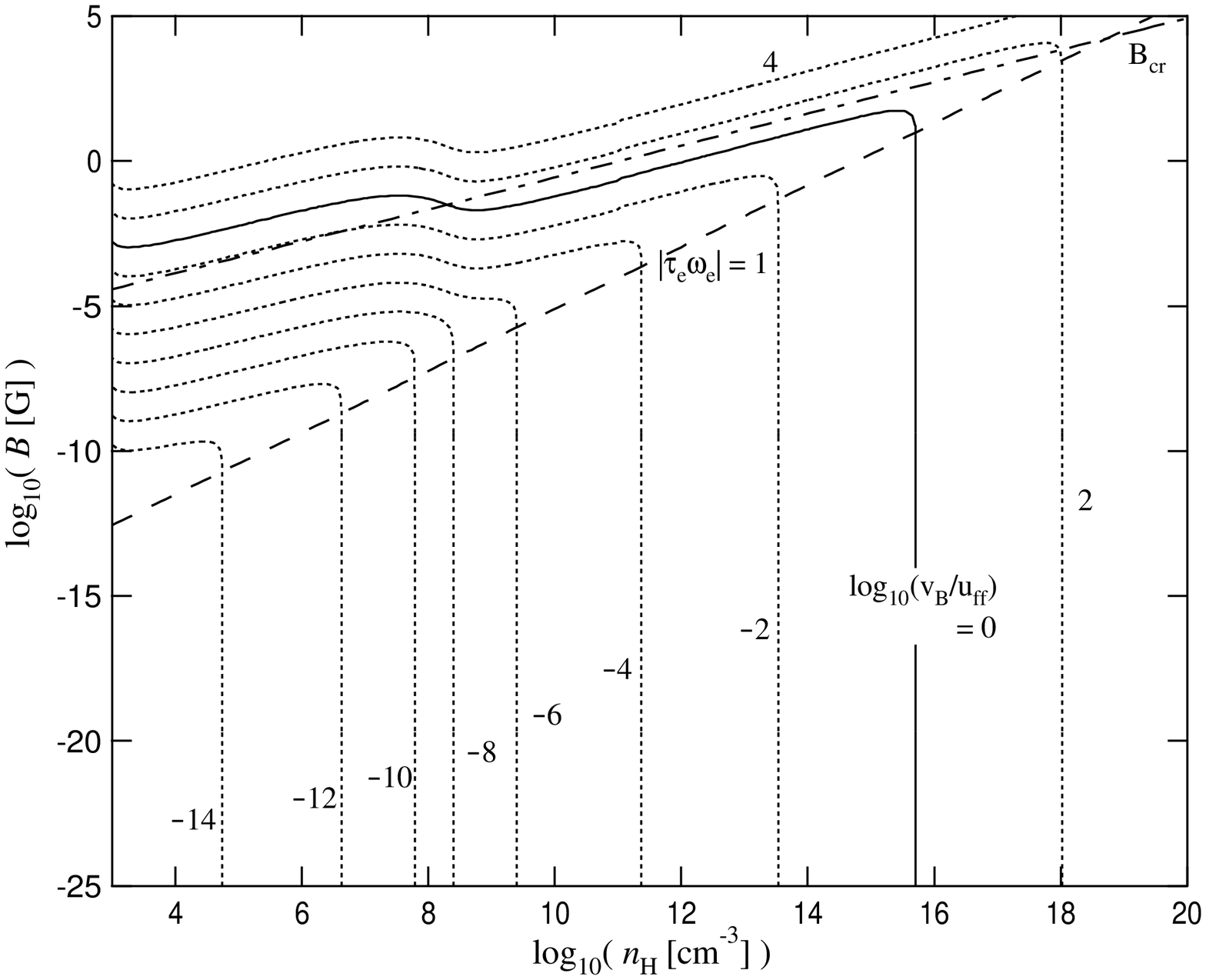}
\caption{Same as Figure \ref{fig:dv_B} except that this model is
 calculated without Li.}
\label{fig:dv_Bc}
\end{figure}

We calculate the drift velocity of magnetic field using
the results of the time dependent
ionization degree in Figures \ref{fig:chem_A}-\ref{fig:chem_Bc}.
The results are shown in Figures \ref{fig:dv_A}-\ref{fig:dv_Bc} . The dotted
curves in Figures \ref{fig:dv_A}-\ref{fig:dv_Bc} show the contours of the
logarithmic value of the drift velocity normalized by the free-fall
velocity, $v_{{\rm B}x}/u_{\rm ff}$ on $n_{\rm H}$-$B$ plane.
The solid curve represents the locus
along which $v_{{\rm B}x}=u_{\rm ff}$ is satisfied.
Here $u_{\rm ff}$ denotes the free-fall velocity.

The condition $|\tau_{\rm e}\omega_{\rm e}|=1$
holds for electrons at the field strength
\begin{equation}
	B_{\rm e} = \frac{c \mu_{\rm en}n_{\rm n}
			\langle\sigma v\rangle_{\rm en}}{e}.
\end{equation}
The field strength $B_{\rm e}$ is also shown by the dashed line.
When the magnetic field is stronger than $B_{\rm e}$, i.e.
$|\tau_{\rm e}\omega_{\rm e}| > 1$, the ambipolar diffusion
is the dominant process for the field dissipation. On the other hand,
when the magnetic field is weaker than $B_{\rm e}$, i.e.
$|\tau_{\rm e}\omega_{\rm e}| < 1$, magnetic flux is lost due to
the Ohmic dissipation.

The dot-dashed line represents the critical field $B_{\rm cr}$.
As we mentioned previously, since we are interested in the collapsing cloud,
our calculations are appropriate in the region where the field
strength satisfies $B < B_{\rm cr}$.
It is clear from Figures \ref{fig:dv_A} and \ref{fig:dv_B} that the frozen-in
condition $v_{{\rm B}x} < u_{\rm ff}$ is almost alway satisfied as far as $B$
is less than the critical field strength $B_{\rm cr}$.  Therefore, the magnetic
field is always frozen to the gas as long as we consider the
collapsing clouds for both of the initial conditions A and B.

Since the field is frozen to the cloud,
the magnetic flux is conserved during the contraction, i.e.
\begin{eqnarray}
	B&=&10^{-5}\left(\frac{B_{\rm ini}}{10^{-11}\ {\rm  G}}\right)
 	\left(\frac{n_{\rm H,ini}}{10^3\ {\rm cm^{-3}}}\right)^{-2/3}\nonumber\\
	&& \times \left(\frac{n_{\rm H}}{10^{12}\ {\rm cm^{-3}}}\right)^{2/3} {\rm G}.
\label{eq:frozen}
\end{eqnarray}
Following the formula of disk radius $r_{\rm d}$ (equation 17) in \citet{TM04},
we can assess the density when the disk is formed:
\begin{equation}
	n_{\rm d}=2.9\times 10^{12}\left(\frac{f_{\rm Kep}}{0.5}\right)^{-6}
	\left(\frac{M}{10\ M_\odot}\right)^{-20/7}{\rm cm^{-3}}.
\label{eq:disk_density}
\end{equation}
Here $M$ denote the total mass within the radius of the disk, $f_{\rm
Kep}$ is the ratio of circular velocity and Keplarian velocity at the
sonic point of the accretion flow. Remark
that $n_{\rm d}$ is not the density of the accretion disk. It is
evaluated by the simple formula $3 M/\left(4\pi r_d^3\mu m_{\rm
p}\right)$, that is the averaged density within the radius $r_{\rm
d}$. Thus, this density could be compared with the density of the core
calculated in our model.
Combining equations (\ref{eq:frozen}) and (\ref{eq:disk_density}), we
obtain the magnetic field strength in the accretion disk:
\begin{eqnarray}
	B&=&2.0\times 10^{-5}\left(\frac{B_{\rm ini}}{10^{-11}\ {\rm  G}}\right)
 	\left(\frac{n_{\rm H,ini}}{10^3\ {\rm cm^{-3}}}\right)^{-2/3} \nonumber\\
	&&\times \left(\frac{f_{\rm Kep}}{0.5}\right)^{-4}
	\left(\frac{M}{10\ M_\odot}\right)^{-40/21} {\rm G}.
\label{eq:disk_b}
\end{eqnarray}

We also evaluated the diffusion velocities in the runs without
Li. Figures \ref{fig:dv_Ac} and \ref{fig:dv_Bc} represents the results
for models A and B. As expected from the results found in the previous
subsection, the diffusion velocity in model A is not so different from the
results with Li, however, in model B, magnetic field is lost by the
Ohmic loss at $n_{\rm H} \gtrsim 10^{16} {\rm cm^{-3}}$. Of course, this is not the
realistic calculation, but we can learn the significance of Li from this results.

\section{Discussion}
\label{discussion}
The condition for which the MRI can grow at the PopIII accretion disk
is investigated by \citet{TB04}.
This condition requires that the MRI
growth timescale is shorter than the diffusion timescale.
Thus, there is the minimum field strength in the disk for the MRI
to be driven.
According to \citet{TB04}, this minimum field strength in the disk
becomes

\begin{eqnarray}
	B&=&1.1\times 10^{-4}{\rm G} \left(\frac{m_*}{10\ M_\odot}\right)^{1/4}
	\left(\frac{T}{10^4\ {\rm K}}\right)^{-3/4}
	\left(\frac{\ln\Lambda}{10}\right)^{1/2}\nonumber\\
	&&\times\left(\frac{\rho_{\rm disk}}
		{5\times 10^{-10}\ {\rm g\;cm^{-3}}}\right)^{1/2}
	\left(\frac{r}{600\ R_\odot}\right)^{-3/4},
\label{eq:discon}
\end{eqnarray}
where $m_*$ is the stellar mass, $r$ is the radius from the cloud center,
$\rho_{\rm disk}$ denotes the density of the accretion disk
and $\ln \Lambda$ is the Coulomb logarithm. Characteristic values of
these parameters are taken from Figure 3 in \citet{TB04}. 

Hence, it is concluded that if the initial field strength in prestellar
core is at
least $\gtrsim 10^{-10}\ {\rm G}$ at $n_{\rm H}=10^3\ {\rm cm^{-3}}$,
the transport of angular momentum could be driven by the turbulence
due to the MRI in the accretion disk.

In fact, there are the models for the generation of
initial seed magnetic field in the universe
\citep{pudritz89,kulsrud97,widrow02,langer03}.
Among these models, \citet{langer03} propose the generation mechanism of
magnetic field based on the radiation force around very luminous objects
such as QSOs.
According to their study, the magnetic field 
$\sim 10^{-11}-10^{-12}\ {\rm G}$ is generated in the inter galactic matter.
Consequently, field strength is amplified to $\sim 10^{-7}-10^{-8}\ {\rm
G}$ when the clouds collapses to $n_{\rm H}=10^3\ {\rm cm^{-3}}$, which
is the initial condition of our analysis.
Thus, in this case, the MRI can be driven and it could be the possible
mechanism of angular momentum transport.
On the other hand, most of the other seed field generation mechanism
predict $B \lesssim 10^{-19}\ {\rm G}$, which is too small to drive MRI.
Thus, MRI may not be important for the formation of very first stars,
since the number of sources which provide anisotropic radiation field
should be very small at the epoch of very first stars.


\section{SUMMARY}
\label{summary}
In this paper, the dissipation of the magnetic field in the collapsing
primordial gas cloud is investigated using a simple analysis. As a result, we find that
that magnetic field is frozen to the gas as far as the
initial field strength satisfies
$ B\lesssim 10^{-5}(n_{\rm H}/10^3~{\rm cm^{-3}})^{0.55}~{\rm G}$.
This condition holds when the magnetic field does not affect the
dynamics of the gravitational contraction.
It is also found that seed magnetic field induced by radiation force is
strong enough to activate MRI in the accretion disk surrounding PopIII
stars. The MRI induced turbulence might play an important role in
transporting the angular momentum in PopIII accretion disk.

\acknowledgments

We are grateful to J. Tan, who made important comments and suggestions on
this paper as a referee.
We thank R. Nishi, K. Omukai, S. Inutsuka, N. Shibazaki, M. Umemura and
A. Ferrara for stimulating discussion. 
We also thank F. Nakamura who provided the data in his calculations.
The analysis has been made with computational facilities 
at Rikkyo University.
We acknowledge Research Grant from Japan Society for the Promotion of
Science (HS 15740122), Rikkyo University Special Fund for Research (HM
\& HS).





\clearpage

\clearpage 

\clearpage 

\clearpage

\clearpage

\clearpage
\clearpage 

\clearpage 


\end{document}